\documentclass[fleqn,twoside]{article}
\usepackage{axodraw}
\usepackage{espcrc2}

\newcommand{\AmS}{{\protect\the\textfont2
  A\kern-.1667em\lower.5ex\hbox{M}\kern-.125emS}}

\title{A natural framework for bi-large neutrino mixing\thanks{Talk presented at the Coral Gables Conference,
Lago Mar Resort, December 16-22, 2003; and at the Fujihara Seminar, ``Neutrino Mass and SeeSaw Mechanism," KEK,
Japan, February 23-25, 2004; OHSTPY-HEP-T-04-007, June 2004}}

\author{Stuart Raby\address[MCSD]{Department of Physics, The Ohio State University,
        174 W. 18th Ave, Columbus, OH 43210, USA \\ \it Email: raby@pacific.mps.ohio-state.edu}}

\begin{document}

\begin{abstract}
In this talk I describe a natural framework for bi-large neutrino mixing within the context of two models -- 1) a
simple generalization of the MSSM and 2) an SO(10) model.   Our starting point is the Frampton, Glashow, Yanagida
[FGY] neutrino mass ansatz which can easily accomodate bi-large neutrino mixing.   The main point of FGY,
however, is to obtain a theory of neutrino masses with only one possible CP violating angle.  They argue that the
sign of the baryon asymmetry of the universe (assuming leptogenesis) is then correlated with CP asymmetries
possibly observable in accelerator experiments.   Unfortunately, there is a fly in the ointment.  It was later
shown by Raidal and Strumia [RS] that there is a sign ambiguity which frustrates the above correlation. We note
that the Raidal-Strumia ambiguity is resolved in our models. \vspace{1pc}
\end{abstract}

\maketitle

\section{Neutrinos : Masses and Mixing Angles}

Let us first summarize the present values of neutrino masses and mixing angles obtained by fitting atmospheric,
solar, reactor and accelerator neutrino oscillation data.   The atmospheric and solar neutrino masses and mixing
angles are given below
\begin{itemize}
\item  $\Delta m^2_{atm} = |m_3^2 - m_2^2| \approx 3 \times 10^{-3} \; {\rm eV}^2$ \subitem $\sin 2\theta_{atm}
\approx 1$
\item  $\Delta m^2_{sol} = |m_2^2 - m_1^2| \approx 7 \times 10^{-5} \; {\rm eV}^2$ \subitem $\sin 2\theta_{sol}
\leq 1$
\end{itemize}
with an approximate mixing matrix given by { \small \begin{eqnarray} \left( \begin{array}{c} \nu_e \\ \nu_\mu \\
\nu_\tau
\end{array} \right) \approx  \left(
\begin{array}{ccc}
c_{sol} &  s_{sol} & 0 \\ -s_{sol}/\sqrt{2} & c_{sol}/\sqrt{2} & 1/\sqrt{2} \\
-s_{sol}/\sqrt{2} & c_{sol}/\sqrt{2} & -1/\sqrt{2} \end{array} \right)  \left(
\begin{array}{c} \nu_1 \\ \nu_2 \\ \nu_3 \end{array} \right) . & & \nonumber
\end{eqnarray}}
For this talk, I will assume three light neutrinos.  Recall, the observed small neutrino masses can naturally be
generated via the See-Saw mechanism.   The $3 \times 3$ neutrino mass matrix is given by $\sim  m_\nu^T \
M_N^{-1} \ m_\nu$, where  $m_\nu$ ($M_N$) is the Dirac RL (Majorana RR) mass matrix.   Of course, the main
problem with neutrino mixing angles is the fact that they are significantly larger than CKM mixing.   These large
mixing angles may be obtained from one ( or a combination ) of the following sources --  $m_\nu$,  $M_N$; or
$m_e$ (the charged lepton mass matrix).  In the models presented below, the origin of large neutrino mixing
angles is in $m_\nu$.

\subsection{Frampton-Glashow-Yanagida ansatz}

Consider now the FGY \cite{Frampton:2002qc} ansatz given by
\begin{eqnarray} {\cal L} & = \ \left( \nu_e \ a \ + \ \nu_\mu \  a^\prime \ e^{-i \phi/2} \right) \ N_1 &
 \\
& + \ \left( \nu_\mu \  b  \ + \ \nu_\tau \  b^\prime \right) \ N_2 &
\nonumber \\
& + \ \frac{1}{2}\ \left(  M_1 \ N_1^2 \ + \ M_2 \ N_2^2 \right) & \nonumber \\ {\cal L} & \equiv \ \nu \ D^T \ N
\ + \ \frac{1}{2} \ N \  M_N \ N  & \nonumber
\end{eqnarray}
where $N_{1,2}$ are the two right-handed (sterile) neutrinos and \begin{eqnarray}
   D^{T} = & \left( \begin{array}{cc} a & 0 \\
a^\prime \ e^{-i \phi/2} & b \\ 0 & b^\prime \end{array} \right) . & \end{eqnarray}   This neutrino mass matrix
ansatz is expressed in the lepton flavor eigenbasis.   The dimensionful parameters $a, \ a^\prime, \ b, \
b^\prime$ are chosen to satisfy the relations
$b \approx b^\prime$ for maximal atmospheric mixing angle and $a \sim a^\prime$ for a large, but not maximal,
solar mixing angle. In addition, due to the zeros in the mass matrix ansatz, there is only one non-vanishing CP
violating angle, $\phi$.
Upon integrating out the heavy sterile neutrinos ($N$), we obtain the $3 \times 3$ FGY light neutrino mass matrix
given by
\begin{eqnarray} {\cal M}_{FGY} = & D^{T} \ M_N^{-1} \ D & \nonumber  \end{eqnarray}
The neutrino mass eigenvalues and small mixing angle are given by
\begin{eqnarray} m_{\nu_3} \approx 2 b^2/M_2 \approx  0.05 \; {\rm eV} =
\sqrt{\Delta m^2_{atm}} & & \nonumber  \\
m_{\nu_2} \approx 2 a^2/M_1 \approx  8.4 \times 10^{-3} \; {\rm eV} =  \sqrt{\Delta m^2_{sol}} & & \nonumber \\
 m_{\nu_1} = 0; \;\;   \theta_{1 3} \sim m_{\nu_2}/(\sqrt{2} \ m_{\nu_3}) & & \nonumber
\end{eqnarray}

\subsection{Raidal and Strumia analysis}

A detailed $\chi^2$ analysis of the FGY ansatz including atmospheric and solar neutrino oscillation data was
performed by Raidal and Strumia \cite{Raidal:2002xf}.  Their best fit then makes predictions for
$\theta_{1 3} = 0.078 \pm 0.015$, observable in planned long baseline experiments, and for the effective electron
neutrino mass measured in neutrinoless double beta decay,  $m_{ee}^{0\nu \beta \beta} = 2.6 \pm 0.4$ meV, which
is unobservable in any planned experiment.

In addition, RS perform a detailed analysis of leptogenesis with FGY.  They find two successful solutions
providing an acceptable cosmological baryon asymmetry [{\em this is the RS ambiguity}].   One with
\begin{eqnarray}  M_1 \, \ll \, M_2; \;\;  M_1 \approx 10^{11} \ {\rm GeV}/|\sin\phi| \; {\rm and} \; \phi < 0 .& &
\end{eqnarray}
In this case, the prediction for CP violating neutrino oscillation is given by
\begin{eqnarray}
 P(\nu_e \rightarrow \nu_\mu) \, < \, P(\nu_\mu \rightarrow \nu_e)  \equiv P(\bar \nu_e \rightarrow \bar
\nu_\mu). & & \nonumber
\end{eqnarray}
In a supersymmetric generalization of FGY, they predict the lepton number flavor violating branching ratios
\begin{eqnarray}  B( \mu \rightarrow e \gamma) \approx  2 \ r \ 10^{-13} ; & & \nonumber \\
 B( \tau \rightarrow \mu \gamma) \geq 3 \ r \ 10^{-12}  & & \nonumber
\end{eqnarray}
where $ r \approx (\tan\beta/10)^2 \ (150 \ {\rm GeV}/ m_{SUSY})^4$.

The other with
\begin{eqnarray}
  M_1 \, \gg \, M_2; \;\; M_2 \approx 10^{12} \ {\rm GeV}/|\sin\phi| \;  {\rm and} \; \phi > 0 . & &
\end{eqnarray}
they find
\begin{eqnarray}
P(\nu_e \rightarrow \nu_\mu) \, > \, P(\nu_\mu \rightarrow \nu_e)  \equiv P(\bar \nu_e \rightarrow \bar \nu_\mu)
& & \nonumber
\end{eqnarray}
and
\begin{eqnarray}  B( \tau \rightarrow \mu \gamma) \approx  7 \ r \ 10^{-11} ; & & \nonumber \\
 B( \mu \rightarrow e \gamma) \geq r \ 10^{-11} . & & \nonumber
\end{eqnarray}

\section{``Natural" FGY texture in SUSY with [SU(2)$\times$U(1)]$_{FS}$}

We now obtain the FGY texture in a supersymmtric theory with an [SU(2)$\times$U(1)]$_{FS}$ family symmetry
protecting the zeros \cite{Raby:2003ay}.  This is a theory of leptons with the 3 families of electroweak doublets
given by $l_i = \left( \begin{array}{c} \nu_i \\  e_i \end{array} \right)$.   The first two families transform as
a doublet under the SU(2) family symmetry with
$L_a =  l_a, \ a = 1,2$,  and then $l_3$ is an SU(2) singlet.
The superpotential for the neutrino mass sector is given by (we will discuss the charged lepton mass sector next)
\begin{eqnarray} W = & \frac{{H_u}}{M} \ ( \ L_a \ \phi^a \ N_1 \ +
 \ L_a \ \tilde \phi^a \ N_2 \  +  \ l_3 \ \omega  \ N_2 \ ) &
 \nonumber \\ & \ + \
\frac{1}{2} \ ( \ S_1 \ {N_1}^2 \ + \ \ S_2 \ {N_2}^2 \ ) &  \end{eqnarray} where $M$ is a new scale satisfying
$M \gg M_Z$.
The familon fields $\phi_a, \ \tilde \phi_a, \ \omega$ are assumed to get the following vacuum expectation values
[VEVs] breaking the family symmetry and generating neutrino masses.  We have, in complete generality, $\langle
\phi \rangle = \left(\begin{array}{c} \langle \phi^1 \rangle \\ \langle \phi^2 \rangle
\end{array}  \right)$ and $\langle \tilde \phi \rangle = \left(\begin{array}{c} 0 \\
\langle \tilde \phi^2 \rangle \end{array} \right)$.  In addition the two right-handed neutrinos  $N_i, \ i= 1,2$
obtain Majorana masses when the SU(2) singlet familon fields get VEVs given by $\langle S_i \rangle = M_i, \;\; i
= 1,2$.

The family symmetry [with specific U(1) charges for all the fields, as discussed in more detail in the paper
\cite{Raby:2003ay}] is sufficient to make this the most general superpotential consistent with the symmetry.
Hence the zeros of the FGY ansatz are obtained, without fine tuning.   When the Higgs doublet $H_u$ gets a VEV at
the weak scale given by
$ \langle H_u \rangle \; = \;
\left(\begin{array}{c} 0 \\
v \sin\beta/\sqrt{2}
\end{array} \right)$
we finally obtain the FGY neutrino mass matrix with the parameters $a, \ a^\prime, \ b, \ b^\prime$ given below.
\begin{eqnarray}  a \;  = \;  v \sin\beta \ \frac{\langle \phi^1 \rangle}{\sqrt{2} M}, & \;\;\; &
a^\prime \ e^{-i \phi/2} \; = \; v \sin\beta \ \frac{\langle \phi^2 \rangle}{\sqrt{2} M}, \nonumber \\
 b \; = \; v \sin\beta \ \frac{\langle \tilde \phi^2 \rangle}{\sqrt{2} M}, & \;\;\; & b^\prime \; =
\; v \sin\beta \ \frac{\langle \omega \rangle}{\sqrt{2} M }.
\end{eqnarray}
Note, however, unlike FGY we must now consider the charged lepton mass matrix which is also constrained by the
family symmetry.   Until we do this, we cannot be certain that we have obtained the FGY ansatz in the lepton
flavor basis.

\section{Charged lepton masses}

Consider the superpotential for charged leptons given by
\begin{eqnarray} W_{ch. \; leptons} = &
\frac{H_d}{M} \ ( \ L_a \ \phi^a \ \bar e_1 \ +
 \ L_a \ \tilde \phi^a  \ \bar e_2 \ +  & \nonumber \\ & \ l_3 \ ( \ \omega \
 \bar e_2
  \ + \ \bar \omega \ \bar e_3 \ )) &   \end{eqnarray}
where the left-handed anti-leptons $\bar e_i, \ i = 1, 2, 3$ are SU(2) singlets. When the Higgs $H_d$ gets a VEV
given by
$ \langle H_d \rangle \; = \; \left( \begin{array}{c}  v \cos\beta/\sqrt{2} \\ 0 \end{array} \right)$ we obtain
the charged lepton $3 \times 3$ mass matrix below

\begin{eqnarray} m_l = \left( \begin{array}{ccc}
\bar a & \bar a^\prime \ e^{-i \phi/2} & 0 \\ 0 & \bar b & \bar b^\prime \\
0 & 0 & \bar c \end{array} \right)  \end{eqnarray} with
\begin{eqnarray}  \bar a \;  = \;  v \cos\beta \
\frac{\langle \phi^1 \rangle}{\sqrt{2} M}, & \;\;\; & \bar a^\prime \ e^{-i \phi/2} \; = \; v \cos\beta \
\frac{\langle \phi^2 \rangle}{\sqrt{2} M}, \nonumber \\  \bar b \; = \; v \cos\beta \ \frac{\langle \tilde \phi^2
\rangle}{\sqrt{2} M}, & \;\;\; & \bar b^\prime \; = \; v \cos\beta \ \frac{\langle \omega \rangle}{\sqrt{2} M},
\nonumber
\\  \bar c = v \cos\beta \ \frac{\langle \bar \omega \rangle}{\sqrt{2} M} & & \nonumber
\end{eqnarray}
satisfying $\bar a, \;\; \bar a^\prime \; \ll \; \bar b, \;\; \bar b^\prime \; \ll \; \bar c$. Note the
parameters $\bar a, \ \bar a^\prime, \ \bar b, \ \bar b^\prime$ are, up to order one coefficients, the same as
$a, \ a^\prime, \ b, \ b^\prime$ appearing in the neutrino mass matrix.

The charged lepton mass eigenvalues are approximately given by $ m_e \ \approx \ \bar a, \; m_\mu \ \approx\ \bar
b, \; m_\tau \ \approx \ \bar c $ and the charged lepton mass matrix is diagonalized by the bi-unitary
transformation $m_l^{diagonal} = U_{\bar e}^\dagger \ m_l \ U_e \ $ with the unitary matrix defining the
left-handed mass eigenstates satisfying $ U_e \approx Diag ( 1, e^{i \phi/2},  -e^{i \phi/2})$.  Using this
matrix we finally obtain the neutrino mass matrix in the flavor basis.   We find
\begin{eqnarray} {\cal M} = & U_e^{T} \; [ \ D^{T} \ M_N^{-1} \ D \ ] \; U_e  \approx {\cal M}_{FGY} &  \end{eqnarray}
with
\begin{eqnarray}   D^{T} = & \left(  \begin{array}{cc} a & 0 \\
a^\prime \ e^{-i \phi/2} & b \\ 0 & b^\prime \end{array} \right).  & \nonumber \end{eqnarray}  Note, the ratios
$|a/a^\prime|$ and $|b/b^\prime|$ can be adjusted to accommodate bi-large neutrino mixing.

As a bonus we now see that the ratio of Majorana neutrino masses $M_1/M_2$ is fixed.   We find
$ (m_e/m_\mu)^2 \approx (\bar a/\bar b)^2 \approx (a/b)^2 \approx (M_1/M_2) (m_{\nu_2}/m_{\nu_3})$. Thus
$(M_1/M_2) \sim 10^{-4}$ and the RS ambiguity is resolved!

\subsection{Related issues}

Note, an SU(2) family symmetry in SUSY theories is desirable for completely different reasons.  It has been shown
that it can ameliorate the SUSY flavor problem \cite{Dine:1993np}.   For example, prior to SU(2) symmetry
breaking the first and second generation sleptons are degenerate.   Hence the off diagonal smuon-selectron mass
term necessary for processes such as $\mu \rightarrow e \gamma$, seen in the figure below, are suppressed.

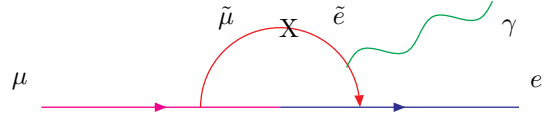
\begin{center}
\SetPFont{Helvetica}{15}

\begin{picture}(300,70)(0,0)

\SetColor{Magenta} \ArrowLine(40,30)(130,30) \Text(35,40)[rc]{$\mu$}

\SetColor{Red}  \LongArrowArcn(130,30)(30,180,0)

\SetColor{Green}  \Photon(155,45)(210,70){4}{2} \Text(220,60)[rc]{$\gamma$}

\SetColor{Black}   \Text(130,60)[lc]{X}  \Text(155,65)[rc]{$\tilde e$} \Text(113,64)[rc]{$\tilde \mu$}

\SetColor{Blue} \ArrowLine(130,30)(220,30)
 \Text(225,40)[lc]{$e$}

\SetColor{Black}
\end{picture}
\end{center}

Finally, at the moment we have an effective higher dimensional field theory with new scales $M \sim M_1 \sim M_2
\gg M_Z$.   Recall that successful leptogenesis requires $\; M_1, \ M_2 \ > \ 10^{11}$ GeV.   Hence it is
reasonable to expect that $ M \ \sim \ M_{GUT}  \approx 3 \times 10^{16}$ GeV.  Consider the following virtues of
SUSY GUTs.   We have
\begin{itemize}

\item $M_{Z} << M_{GUT}$ ``Naturally"

\item Explains Charge Quantization

\item Predicts Gauge Coupling Unification$^*$

\item Predicts Yukawa Coupling Unification

\item  + Family Symmetry $\Longrightarrow$ Hierarchy of Fermion Masses and Protects against large flavor
violation

\item Neutrino Masses via See - Saw scale $\sim 10^{-3} - 10^{-2} \ M_G$  $\sim M_G^2/M_{Pl}$

\item LSP -- Dark Matter Candidate, and

\item Baryogenesis via Leptogenesis
\end{itemize}
With all of these virtues it is worth considering embedding our generalized MSSM into an SO(10) SUSY GUT.  We
consider the following SO(10) SUSY GUT with an $[SU(2) \times U(1)^n]_{FS}$ proposed intially in
\cite{Barbieri:1996ww} and analyzed in great detail in \cite{Blazek:1999ue}.

\section{SO(10) SUSY GUT $\times$ $[SU(2) \times U(1)^n]_{FS}$}

The superpotential for the charged fermion sector, including the heavy Froggatt-Nielsen [FG] states \{ $\chi^a$,
\ $\bar \chi_a$ \}, familon fields \newline \{ $\phi^a, \ \tilde \phi^a$ \} and SO(10) adjoint \{ 45 \}, is given
by
\begin{eqnarray} W \supset & 16_3 \ 10_H \ 16_3 +  16_a \ 10_H \ \chi^a &  \\
 & +  \bar \chi_a \ (M_{\chi} \ \chi^a + \ 45 \ \frac{\phi^a}{ M} \  16_3  \nonumber \\
 & + \ 45 \ \frac{\tilde \phi^a \tilde \phi^b}{ M^2} \  16_b + A^{a b} \ 16_b ) & \nonumber
\end{eqnarray}
where we have the three families in $16_a \;  (a, b = 1, 2), \ 16_3$; and the Higgs doublets in $10_H$.

In addition, the FG mass $M_\chi$ necessarily includes SO(10) breaking VEVs  with $M_\chi = M (1\;\;+ \;\; \alpha
\; X \;\; +\;\; \beta \; Y)$. $ X,\; Y$ are the SO(10) breaking VEVs in the adjoint representation of SO(10) with
$X$ corresponding to the  U(1) in SO(10) which preserves SU(5), and $Y$ the standard weak hypercharge. $\alpha,
\; \beta$ are arbitrary parameters.   Upon integrating out the FG fields we obtain the effective fermion mass
operators in the figure.
Finally upon giving the familon fields VEVs we obtain the effective Yukawa couplings below.
\begin{eqnarray}
Y_u =&  \left(\begin{array}{ccc}  0 & \epsilon' \ \rho & - \epsilon \ \xi  \\
             - \epsilon' \ \rho &  \tilde \epsilon \ \rho & - \epsilon    \\
       \epsilon \ \xi   & \epsilon & 1 \end{array} \right) \; \lambda & \nonumber \\
Y_d =&  \left(\begin{array}{ccc}  0 & \epsilon'  & - \epsilon \ \xi \ \sigma \\
- \epsilon'   &  \tilde \epsilon  & - \epsilon \ \sigma \\
\epsilon \ \xi  & \epsilon & 1 \end{array} \right) \; \lambda & \label{eq:yukawa} \\
Y_e =&  \left(\begin{array}{ccc}  0 & - \epsilon' & 3 \ \epsilon \ \xi \\
          \epsilon'  &  3 \ \tilde \epsilon  & 3 \ \epsilon  \\
 - 3 \ \epsilon \ \xi \ \sigma  & - 3 \ \epsilon \ \sigma & 1 \end{array} \right) \; \lambda &
 \nonumber \end{eqnarray}
with  \begin{eqnarray}  \xi \;\; =  \;\; \langle \phi^1 \rangle/\langle \phi^2 \rangle ; & \;\;
\tilde \epsilon  \;\; \propto   \;\; (\langle \tilde \phi^2 \rangle/ M)^2 ;  & \nonumber \label{eq:omega} \\
\epsilon \;\; \propto  \;\; \langle \phi^2 \rangle/ M ; &  \;\;
\epsilon^\prime \;\; \sim  \;\;  \langle A^{1 2} \rangle/ M ; \nonumber \\
  \sigma \;\; =   \;\; \frac{1+\alpha}{1-3\alpha}; &  \;\; \rho \;\; \sim   \;\;
  \beta \ll \alpha .& \nonumber
\end{eqnarray}
The small parameters are given by $\epsilon, \ \tilde \epsilon, \ \epsilon^\prime$; all other parameters are
numbers of order one.  All these parameters are fit to the low energy data via a detailed $\chi^2$ analysis.
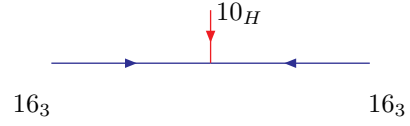
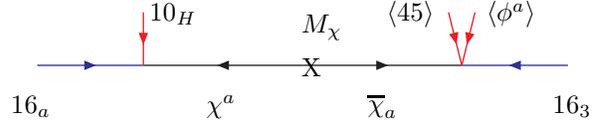
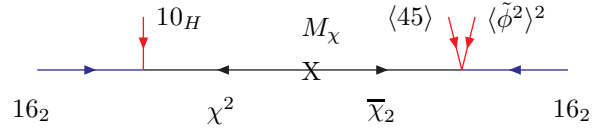
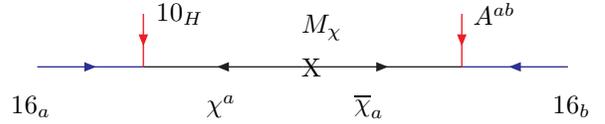
\begin{center}
\SetPFont{Helvetica}{15}

\begin{picture}(200,300)(0,0)

\SetColor{Blue} \ArrowLine(40,280)(100,280) \Text(40,265)[rc]{$16_3$}

\SetColor{Red}  \ArrowLine(100,300)(100,280) \Text(102.5,300)[lc]{$10_H$}

\SetColor{Blue} \ArrowLine(160,280)(100,280) \Text(160,265)[lc]{$16_3$}

\SetColor{Blue} \ArrowLine(0,200)(40,200) \Text(5,185)[rc]{$16_a$}

\SetColor{Blue} \ArrowLine(200,200)(160,200) \Text(195,185)[lc]{$16_3$}

\SetColor{Black} \ArrowLine(100,200)(40,200) \Text(75,185)[rc]{$\chi^a$}

\SetColor{Black} \ArrowLine(100,200)(160,200) \Text(125,185)[lc]{$\overline{\chi}_a$}

\SetColor{Red}  \ArrowLine(40,220)(40,200) \Text(42.5,220)[lc]{$10_H$}

\SetColor{Red}   \ArrowLine(155,220)(160,200) \ArrowLine(165,220)(160,200) \Text(150,220)[rc]{$\langle 45
\rangle$} \Text(170,220)[lc]{$\langle \phi^a \rangle$}

\SetColor{Black} \Text(100,200)[lc]{X}

\SetColor{Black} \Text(100,215)[lc]{$M_\chi$}

\SetColor{Blue} \ArrowLine(0,110)(40,110) \Text(5,95)[rc]{$16_2$}

\SetColor{Blue} \ArrowLine(200,110)(160,110) \Text(195,95)[lc]{$16_2$}

\SetColor{Black} \ArrowLine(100,110)(40,110) \Text(75,95)[rc]{$\chi^2$}

\SetColor{Black} \ArrowLine(100,110)(160,110) \Text(125,95)[lc]{$\overline{\chi}_2$}

\SetColor{Red}  \ArrowLine(40,130)(40,110) \Text(45,130)[lc]{$10_H$}

\SetColor{Red}  \ArrowLine(155,130)(160,110) \ArrowLine(165,130)(160,110) \Text(150,130)[rc]{$\langle 45
\rangle$} \Text(170,130)[lc]{$\langle \tilde \phi^2 \rangle^2$}

\SetColor{Black} \Text(100,110)[lc]{X}

\SetColor{Black} \Text(100,125)[lc]{$M_\chi$}

\SetColor{Blue} \ArrowLine(0,15)(40,15) \Text(5,0)[rc]{$16_a$}

\SetColor{Blue} \ArrowLine(200,15)(160,15) \Text(195,0)[lc]{$16_b$}

\SetColor{Black} \ArrowLine(100,15)(40,15) \Text(75,0)[rc]{$\chi^a$}

\SetColor{Black} \ArrowLine(100,15)(160,15) \Text(120,0)[lc]{$\overline{\chi}_a$}

\SetColor{Red}  \ArrowLine(40,35)(40,15) \Text(45,35)[lc]{$10_H$}

\SetColor{Red}  \ArrowLine(160,35)(160,15) \Text(165,35)[lc]{$A^{a b}$}

\SetColor{Black} \Text(100,15)[lc]{X}

\SetColor{Black} \Text(100,30)[lc]{$M_\chi$}

\SetColor{Black}
\end{picture}
\end{center}

\subsection{Features of the model }
The model has the following nice features.  The family hierarchy is obtained via the hierarchy of family symmetry
breaking $SU_2 \times U_1 \;\;\; \longrightarrow \;\;\; U_1 \;\;\; \longrightarrow \;\;\; {\rm nothing} $ where
the first (second) breaking is due to the familon VEVs determining the small ratios $\epsilon, \ \tilde \epsilon$
($\epsilon^\prime$).   As a result the 3rd family is much heavier than the 2nd family; is then heavier than the
1st family.

We incorporate the following mass patterns -- 1) an approximate Georgi - Jarlskog mechanism ``naturally" with the
familon VEV $\langle 45 \rangle = (B - L) M_G $ giving  $ m_s \sim  \frac{1}{3} m_\mu$ and  $m_d \sim  3 m_e $ ;
2) third generation Yukawa unification with $\lambda_t = \lambda_b = \lambda_\tau = \lambda_{\nu_\tau}$ $=
\lambda$ $@ M_G$ and 3) $\beta \ll \alpha \sim 1 $ leads to $m_u < m_d$ even though $m_t \gg m_b$.

We ``naturally" satisfy gauge coupling unification and the $SU_2$ family symmetry suppresses flavor violation
such as $\mu \rightarrow e \gamma$.

The model has 10 Yukawa parameters (6 real coefficients and 4 phases) to fit 13 fermion masses and mixing angles.
Varying these 10 Yukawa parameters [three gauge ($\alpha_G, \ M_G, \ \epsilon_3$) and 7 soft SUSY breaking
parameters] at $M_G$ we use the two (one) loop RG equations for dimensionless (dimensionful) variables to obtain
the $\chi^2$ function at $M_Z$.   All observables at $M_Z$ are evaluated including one loop threshold
corrections.   The result of the $\chi^2$ analysis, using the code of T. Blazek (see Refs.
\cite{Blazek:1996yv,Blazek:1999ue}), is given in the Table.  The fit is quite good. In particular there are only
10 Yukawa parameters with 16 independent fermion mass and mixing angle observables.
\begin{table*}[htb]
\caption{$\chi^2$ analysis including 5 precision electroweak and 16 fermion mass and mixing angle observables
plus the branching ratio for the process $b \rightarrow s \gamma$.}
$$
\begin{array}{lccc}
\hline
{\rm Observable}  &{\rm Data}(\sigma) & Theory & Pull  \\
\mbox{ }   & {\rm (masses} & {\rm in\  \ GeV) } &  \\
\hline
\;\;\;M_Z            &  91.188 \ (0.091)  &  91.21   & <0.50       \\
\;\;\;M_W             &  80.419 \ (0.080)    &  80.40    & <0.50    \\
\;\;\;G_{\mu}\cdot 10^5   &  1.1664 \ (0.0012) &  1.166   & <0.50   \\
\;\;\;\alpha_{EM}^{-1} &  137.04 \ (0.14)  &  137.0    &  <0.50          \\
\;\;\;\alpha_s(M_Z)    &  0.11720 \ (0.002)   &   0.1139 &   2.65       \\
\hline
\;\;\;M_t              &   174.30 \ (5.1)   &  171.3   & <0.50     \\
\;\;\;m_b(M_b)          &    4.220 \ (0.09)  &    4.377   &  3.04               \\
\;\;\;M_b - M_c        &    3.400 \ (0.2)   &    3.430 &      < 0.50            \\
\;\;\; m_c(m_c)     &   1.3000  \ (0.15)  &   1.212  &       <0.50   \\
\;\;\;m_s              &  0.089 \ (0.011)   &  0.100 & 1.01          \\
\;\;\;m_d/m_s          &  0.050 \ (0.015)   &  0.0751 &  2.80       \\
\;\;\;Q^{-2}           &  0.00203 \ (0.00020)  &  0.00200 & < 0.50                \\
\;\;\;M_{\tau}         &  1.777 \ (0.0018)   &  1.777 & < 0.50         \\
\;\;\;M_{\mu}          & 0.10566 \ (0.00011)   & .1057  & < 0.50         \\
\;\;\;M_e \cdot 10^3      &  0.5110 \ (0.00051) &  0.5110  & < 0.50 \\
\;\;\;V_{us}         &  0.2230 \ (0.0040)      &  0.2213    &  < 0.50   \\
\;\;\;V_{cb}         & 0.04020 \ (0.0019)      &  0.0391  &  < 0.50           \\
\;\;\;V_{ub}/V_{cb}    &  0.0860 \ (0.008)    &  0.0850    & < 0.50             \\
\;\;\;  V_{td} &    0.00820  \  (0.00082) &   0.00846 &    <0.50    \\
\;\;\; \epsilon_K          & 0.00228 \ (0.00023)    &  0.00233 &    < 0.50         \\
\;\;\; \sin2\beta &  0.7270 \ (0.036) &   0.6898 &  1.07     \\
\hline {B(b\!\rightarrow\! s \gamma)\!\cdot\!10^{4}}  &  3.340 \ (0.38) &  3.433 &  < 0.50  \\
\hline
  \multicolumn{2}{l}{{\rm TOTAL}\;\;\;\; \chi^2}  12.16
            & &\\
\hline
\end{array}
$$
\end{table*}

\subsection{Bi-large neutrino mixing in \\ $SO_{10} \times [SU_2 \times U_1^n]_{FS}$ model}
Let us now consider neutrino masses in this model.   The model has three right-handed neutrinos contained in the
fields $16_a, \ 16_3$.   Moreover the Dirac neutrino Yukawa matrix is given by
\begin{eqnarray}
Y_{\nu} =&  \left(\begin{array}{ccc}  0 & - \epsilon' \ \omega & {3 \over 2} \ \epsilon \ \xi \ \omega  \\
      \epsilon'  \ \omega &  3 \ \tilde \epsilon \  \omega & {3 \over 2} \ \epsilon \ \omega \\
       - 3 \ \epsilon \ \xi \ \sigma   & - 3 \ \epsilon \ \sigma & 1 \end{array} \right) \; \lambda &
 \end{eqnarray} with $\omega \;\; =  \;\; 2 \, \sigma/( 2 \, \sigma - 1)$
This then gives the Dirac LR neutrino mass matrix --  $ m_\nu \equiv Y_\nu \frac{v}{\sqrt{2}} \sin\beta $.

We now add three SO(10) singlet fields  $N_i, \ i = 1,2,3$ and couple them to neutrinos via the superpotential
\begin{eqnarray} W_{neutrino} = & \frac{\overline{16}}{ M}  \left( N_1 \ \tilde \phi^a \ 16_a \ + \ N_2 \
\phi^a \ 16_a \right) \nonumber \\ & + \frac{\overline{16}}{ M}  \left( \ N_3 \ \theta \ 16_3 \right) &   \\ & +
\frac{1}{2} \left( S_1 \ N_1^2 \ + \ S_2 \ N_2^2 \right) .  & \nonumber
\end{eqnarray}

The field $\overline{16}$ (and another field $16$) are assumed to obtain VEVs in the right-handed neutrino
direction.  In fact, these VEVs are necessary (along with the $45$ VEVs) to break SO(10) to the standard model.
Given $\langle \overline{16} \rangle = V$, $\langle S_i \rangle = M_i, \ i = 1,2$ and the familon VEVs we obtain
the effective neutrino mass terms given by
\begin{eqnarray} W^{eff}_{\nu} = & \nu \ m_\nu \ \bar \nu + \bar \nu \ V \ N +
\frac{1}{2} \ N \ M_N \ N  &  \end{eqnarray} where
\begin{eqnarray} (V^{T})^{-1} = & \frac{ M}{v_{16}} \left( \begin{array}{ccc}
 - 1/ (\langle \tilde \phi^2 \rangle \ \xi) & 1/ \langle \phi^1 \rangle  & 0 \\
1/ \langle \tilde \phi^2 \rangle  & 0 & 0 \\
0 & 0 & 1/ \langle \theta \rangle
\end{array} \right) . & \nonumber
\end{eqnarray}

Once the heavy $\bar \nu, \ N$ fields are integrated out near the GUT scale, we obtain the $ 3 \times 3$ Majorana
neutrino mass matrix (in the flavor basis)
\begin{eqnarray} {\cal M} = & U_e^{T} \; [ \ m_\nu \ (V^{T})^{-1} \ M_N  \ V^{-1} \
m_\nu^{T} \ ] \; U_e . &
\end{eqnarray}

Note, defining the $3 \times 2$ Dirac neutrino mass matrix
\begin{eqnarray}  D^{T}  \equiv & m_\nu \
(V^{T})^{-1} \ M_N \ \left( \begin{array}{cc} 1 & 0 \\ 0 & 1 \\ 0 & 0 \end{array} \right) = \left(
\begin{array}{cc}
a & 0  \\
a^\prime  & b  \\
0 & b^\prime
\end{array} \right)  & \nonumber
\end{eqnarray}
and the $2 \times 2$ right-handed neutrino mass matrix
\begin{eqnarray}  & {\hat M_N} \equiv \left( \begin{array}{cc} M_1 & 0 \\ 0 & M_2 \end{array} \right), & \nonumber
\end{eqnarray} we finally obtain the light neutrino mass matrix
\begin{equation} {\cal M} = U_e^{T} \; [ \ D^{T} \ \hat M_N^{-1} \ D \ ] \; U_e . \end{equation}
Hence we obtain the Frampton, Glashow \& Yanagida ansatz without any fine-tuning.  In particular, the zeros are
exact and fixed by the family symmetry.  This may be seen by the intermediate calculation below.

We have
\begin{eqnarray}
& Y_{\nu} \;\; (V^{T})^{-1}  &  \\
& \sim \left(\begin{array}{ccc}  0 & - \epsilon' \ \omega & {3 \over 2} \ \epsilon \ \xi \ \omega  \\
      \epsilon'  \ \omega &  3 \ \tilde \epsilon \  \omega & {3 \over 2} \ \epsilon \ \omega \\
       - 3 \ \epsilon \ \xi \ \sigma   & - 3 \ \epsilon \ \sigma & 1 \end{array} \right) \;  &   \nonumber
       \\
& \times \; \left( \begin{array}{ccc}
 - 1/ (\langle \tilde \phi^2 \rangle \ \xi) & 1/ \langle \phi^1 \rangle  & 0 \\
1/ \langle \tilde \phi^2 \rangle  & 0 & 0 \\
0 & 0 & 1/ \langle \theta \rangle
\end{array} \right)  & \nonumber \\
& \sim \left(
\begin{array}{ccc}
a & 0  & {\rm X} \\
a^\prime  & b & {\rm X} \\
0 & b^\prime & {\rm X}
\end{array} \right)   & \nonumber
 \end{eqnarray}

In addition,  bi-large mixing requires significant constraints on the parameters $b, \ b^\prime, \ a, \
a^\prime$.    Maximal atmospheric neutrino oscillation requires $b \; \approx \; b^\prime$, whereas large mixing
for solar neutrino oscillations requires  $a$ of order $a^\prime$.   On the other hand, the ratios $|b/b^\prime|$
and $|a/a^\prime|$ are, now, also constrained by charged fermion masses and mixing angles.  Without performing a
joint $\chi^2$ analysis (including both charged fermions and neutrinos) we can still make the following
observations. The fit to charged fermion masses naturally gives $\epsilon^\prime \ \sim \ \epsilon \ \xi$. Hence
given
\begin{eqnarray} b \equiv &
\epsilon^\prime \ \omega \ \lambda \ (M_2/ \phi^1) \
\frac{ M}{ v_{16}} \frac{v \sin\beta}{\sqrt{2}} &  \\
b^\prime \equiv & - 3 \ \epsilon \ \xi \ \sigma \ \lambda \ (M_2/ \phi^1) \ \frac{ M }{ v_{16}} \frac{v
\sin\beta}{\sqrt{2}} . & \nonumber
\end{eqnarray}
we find $|b/b^\prime| = (\epsilon^\prime \ \omega)/( 3 \ \epsilon \ \xi \ \sigma)  \approx 1$.   With regards $a,
\ a^\prime$ we have
$\epsilon^\prime \ \xi^{-1} \ \sim \ \tilde \epsilon$.   Moreover
\begin{eqnarray} a \equiv & - \epsilon^\prime \ \omega \ \lambda \ (M_1/
\tilde \phi^2) \ \frac{ M }{ v_{16}} \frac{v
\sin\beta}{\sqrt{2}} &  \\
a^\prime \equiv & ( -\epsilon^\prime \ \xi^{-1} + 3 \ \tilde \epsilon ) \ \omega \ \lambda \ (M_1/ \tilde \phi^2)
\ \frac{ M }{v_{16}} \frac{v \sin\beta}{\sqrt{2}} & \nonumber \end{eqnarray}
gives $|a/a^\prime| = \epsilon^\prime/(( -\epsilon^\prime \ \xi^{-1} + 3 \ \tilde \epsilon )$ which apparently
requires fine-tuning of order 1 in 10 to get $a \sim a^\prime$.   Clearly in order to test this model, we must
perform an extended $\chi^2$ analysis including both charged fermions and neutrinos.  Note, there are only two
new parameters in the neutrino sector which can be fit to the solar and atmospheric neutrino mass squared
differences.

Finally we find $m_{\nu_2}/m_{\nu_3} \approx (m_e/m_\mu) \ (M_1/M_2) \ \tilde \epsilon $ and thus $M_1/M_2 \sim
10^3$.   Unfortunately, leptogenesis is more complicated in this model since we must first identify the heavy
neutrino mass eigenstates and their lepton/Higgs couplings.  Moreover there are, in principle, more CP violating
phases.

\subsection{Summary}

The Frampton-Glashow-Yanagida neutrino mass matrix ansatz has several nice features.   Since it has only one CP
violating angle it can, in principle, correlate the sign of the matter - anti-matter asymmetry with CP violating
neutrino oscillations.   Unfortunately, the Raidal-Strumia analysis uncovered an ambiguity which allows either
sign of the CP violating angle for successful leptogenesis, depending on whether the right-handed neutrino masses
satisfies  $M_1 \ll M_2$ or $M_1 \gg M_2$!  In addition the FGY ansatz provides a natural framework for bi-large
neutrino mixing!

In this talk we obtained the FGY ansatz in a supersymmetric model with an [SU(2) $\times$ U(1)]$_{FS}$ family
symmetry.  This had two advantages, 1) the FGY ansatz is fixed by the family symmetry and 2)  it resolves the RS
ambiguity.   This is because the same family symmetry acting on neutrinos also constrains the charged fermion
masses.  As a result,  the ratio of right-handed neutrino masses is fixed by the ratio of $\delta m^2$ for
atmospheric and solar neutrinos and the ratio of charged fermion masses.    It is also important to note that we
are only able to obtain the normal hierarchy for neutrino masses.

The second model presented in this talk was a SUSY SO(10) model with an [SU(2) $\times$ U(1)$^n$]$_{FS}$ family
symmetry.   This model is significantly more constrained since it relates all fermion masses and mixing angles.
There are only two new arbitrary parameters in the neutrino sector which are fixed by the neutrino mass
differences.   Nevertheless, we ``naturally" obtain the FGY ansatz for neutrinos (in a theory which initially
starts with 3 right-handed neutrinos and 3 SO(10) singlet neutrinos). We find a large neutrino mixing angle for
atmospheric neutrino oscillations ``naturally."  On the other hand, an extended $\chi^2$ analysis must be
performed to see if we can obtain a large solar mixing angle.   If a good fit is obtained for neutrino masses and
mixing, it will then be very interesting to analyze CP violation and leptogenesis in this model.

\noindent
{\bf Acknowledgements}

This work is partially supported by DOE/ER/01545-857.

\end{document}